# Bubbles dissolution in cylindrical microchannels

**Rivero-Rodriguez, Javier**† **and Scheid, Benoit.**

[1]TIPs, Université Libre de Bruxelles, C.P. 165/67, Avenue F. D. Roosevelt 50, 1050 Bruxelles, Belgium

This work focuses on the dissolution of a train of unconfined bubbles in cylindrical microchannels. We investigate how the mass transfer is affected by the channel and bubble sizes, distance between bubbles, diffusivity, superficial velocity, deformation of the bubble, the presence of surfactants in the limit of rigid interface and off-centered positions of the bubbles. We analyse the influence of the dimensionless numbers and especially the distance between bubbles and the Péclet number, $Pe$, which we vary among eight decades and identify five different dissolution regimes. We show different concentration patterns and the dependence of the Sherwood numbers, $Sh$. These regimes can be classified by either the importance of the streamline diffusion or by the interaction between bubbles. For small $Pe$ the streamline diffusion is not negligible as compared to convection whereas for large $Pe$, convection dominates in the streamline direction and, thus, crosswind diffusion becomes crucial in governing the dissolution through boundary layers or of the remaining wake behind the bubbles. Bubbles interaction takes place for very small $Pe$ for which the dissolution is purely diffusive or for very large $Pe$ numbers in which case long wakes eventually reach the following bubble. We also observe that the bubble deformability mainly affects the $Sh$ in the regime for very large $Pe$ in which bubbles interaction matters, and also that the rigid interface effect affects the boundary layer and the remaining wake. The effect of off-centered position of the bubble, determined by the transverse force balance, is also limited to large $Pe$. The boundary layers in rigid bubble surfaces are thicker as compared to those on stress-free bubble surface and, thus, the dissolution is smaller. For centered bubbles, the influence of inertia on the dissolution is negligible. Finally, we discuss underlying hypothesis of the model such as the truncated modal expansion as well as the quasi-steadiness and periodicity assumptions. We have carried out the numerical simulations based on the finite element method, solving the steady Stokes equations for the hydrodynamics and the eigenvalue counterpart of the unsteady advection-diffusion equation after separation of variables for the mass transfer.

## 1. Introduction

Nowadays, microfluidics is undergoing rapid increase due to a deeper control of operating conditions and the possibility of continuous production. Microfluidic devices are present in a wide range of applications, examples of them being from dissolution of $CO_2$ in microchannels (Cubaud & Ho 2004; Chi *et al.* 2017; Durgadevi & Pushpavanam 2018), to culture of cells (Mehta *et al.* 2007), adsorption kinetics of surfactants (Riechers *et al.* 2016), polymerisation of monodisperse beads by solvent evaporation in microfluidics (Hung *et al.* 2010) or boiling and condensation in microchannels (Kandlikar *et al.* 2005). Despite the apparent lack of relation between these examples, they have important similarities. In effect, objects such as bubbles, cells or drops are immersed in a liquid flowing in small microchannels and there exists an exchange of mass/heat between these objects and the surrounding liquid. Beside microfluidics, other phenomena also rely on mass/heat transfer in microchannels such as microrocket production (Gallino *et al.* 2018), oxygen transport by red blood cells in capillaries (Vadapalli *et al.* 2002) and the control of the nucleation and full dissolution of bubbles in extracorporeal medical devices, whose presence in blood may leads to gas embolism (Barak & Katz 2005). Thus, it is high importance to control the transfer mechanisms governing all these processes.

Kuo & Chiu (2011) provide a review of available techniques for the control of mass transport in microfluidics. Most of them rely on very sophisticated geometries or on the tailoring the surface rheology Beltramo *et al.* (2017). In this work, we consider the mass transfer of bubbles in cylindrical microchannels, i.e. straight microchannels with a circular cross-section and study the influence of all the properties of the system in a systematic manner using numerical means. To understand this problem, the hydrodynamics and the dissolution are studied in two different steps. The hydrodynamics is reported in a previous work by the authors Rivero-Rodriguez & Scheid (2018).

† Email address for correspondence: jriveror@ulb.ac.be

The dissolution of $CO_2$ in water flowing in microchannels has recently drawn the attention of many researchers (Cubaud *et al.* 2012; Shim *et al.* 2014; Ganapathy *et al.* 2014; Mikaelian *et al.* 2015*b*). Despite the fact that larger scale systems, such as column reactors are studied for several decades (Griffith 1960; Deckwer 1980), it is still a matter of active research because of its rich dynamics (Auguste & Magnaudet 2018) and dissolution regimes (Dhotre & Joshi 2004). The miniaturisation of bubble columns is known to increase the mass transfer for both gas-liquid (Durgadevi & Pushpavanam 2018) and liquid-liquid (Xu *et al.* 2008) situations, which increases the reaction rate in mass transfer limited reactors (Kashid *et al.* 2011). In the case of bubbles in microchannels, few hydrodynamics regimes has been found (Cubaud & Ho 2004; Abiev 2013) relative to the size and ordering of the bubbles, or relative to coalescence (Cubaud & Ho 2004). Concerning the dissolution of a train of bubbles, the pure diffusion has been considered by Shim *et al.* (2014) as well as its dissolution through a boundary layer (Mikaelian *et al.* 2015*a*). However, there is a lack in the literature of the characterisation of all the different dissolution regimes. In this work, we provide a complete picture of the dissolution of bubbles in microchannels covering the entire range of experimentally relevant distance between bubbles and Péclet numbers from small (Shim *et al.* 2014) to large (Cubaud *et al.* 2012). It is useful for the control of this phenomenon in many applications such as reactions in bubbly flow (Hashemi *et al.* 2015) or controling the oxygen level of cells (Mehta *et al.* 2007). Modelisation of similar situations has been already carried out for the control of glucose concentration in a culture of cells (Westerwalbesloh *et al.* 2015) or for the design of biosensors (Squires *et al.* 2008).

The structure of the paper is as follows. In sec. 2 we present the model we use to describe the dynamics and dissolution of the train of bubbles of finite size and explain the underlying hypotheses. In sec. 2.1 we set up the equations governing the hydrodynamics and in sec. 2.2 we derive the equations governing the dissolution. Additional comments on the scaling and the numerics are given in sec. 2.3. In sec. 3 we study the effect of all the parameters on the dissolution of bubbles under the hypotheses of stress-free and spherical interface, creeping flow and centered bubbles. We provide a qualitative description of the different dissolution regimes and characterise it depending on the importance of streamline diffusion. On the one hand, the behaviour for small $Pe$ numbers, implying that the streamline diffusion is not negligible as compared to convection, is presented in sec. 3.1 including a simplified model. On the other hand, the behaviour for large $Pe$ is presented in sec. 3.2. In this case, convection dominates along the streamline direction and, thus, crosswind diffusion becomes crucial to govern the dissolution through a boundary layer or of the remaining wake behind the bubbles. We also studied the local effects in the limit of small distance between bubbles in sec. 3.3. Next, we consider some extensions of the previous model in sec. 4, such as the limit of a rigid interface in sec. 4.1, the deformability in sec. 4.2, the inertial forces in sec. 4.3 and off-centered positions in sec. 4.4. The validity of the hypotheses is discussed in sec. 5. Finally, conclusions and discussion are presented in sec. 6.

## 2. Modelling

We study the dissolution of a train of bubbles in a cylindrical microchannel by numerical means. To model this situation, we solve the steady Stokes equations and the advection-diffusion equation among a periodic volume $\mathcal{V}$ of length $L$ containing one single bubble of volume $\mathcal{V}_B$, where $d$ is the bubble diameter, attached to it and delimited by the walls of the channel, $\Sigma_W$, two cross sections of the channel, $\Sigma_{IN}$ and $\Sigma_{OUT}$, and the surface of the bubble, $\Sigma_B$, as schematised in fig. 1. Equivalent diameters are defined as referred to the volume, $\mathcal{V}_B = \frac{\pi}{6}d^3$ or to the surface, $\Sigma_B = \pi d_\Sigma^2$. A liquid of constant properties, such as viscosity $\mu$, surface tension $\gamma$, isotropic gas diffusivity in the liquid $D$ and saturation concentration of the gas in the liquid $C_{sat}$, flows inside a channel of diameter $d_h$ with a superficial velocity $J$ producing a pressure drop, $\Delta p$, due to the Poiseuille flow modified by the presence of one bubble. The bubble travels at the velocity $V$, which is determined by the balance of forces acting on the bubble surface in the axial direction, and dissolves with a mass transfer coefficient $k_\ell$, which is determined by the slowest eigenmode, as explained below. Consequently, both values, $V$ and $k_\ell$, are computationally determined.

It is assumed that the characteristic time in which change of volume of the bubble occurs, either due to dissolution or due to the compressibility of the gas, is large as compared to the time a bubble travels its own size, namely the hydrodynamic time. Under this assumption, the flow field can be considered as quasi-steady in absence of vortex shedding and turbulence, as it usually happens in microfluidics and considered in this work. The validity of the periodicity and quasi-steadiness hypotheses will be discussed after presenting the results. First, we consider a stress-free and non-deformable spherical interface, in the case of a creeping flow and centered bubbles. Later, we extend the model to consider a rigid interface, deformable interface, inertial forces and off-centered bubbles.

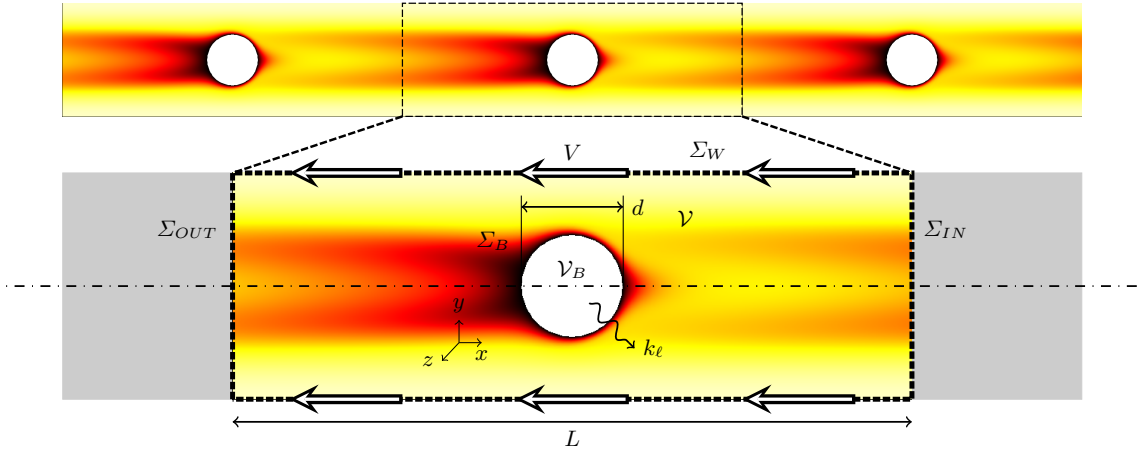

FIGURE 1. Sketch of the modelled segment of a train of bubbles in a microchannel.

### 2.1. *Modelling the hydrodynamics*

According to the previous description, the liquid flow around the bubble in a channel segment of length $L$, being $L$ the distance between two consecutive bubbles from center to center, can be analysed by solving the continuity and the steady Stokes equations,

$$\boldsymbol{\nabla} \cdot \boldsymbol{v} = 0 \qquad \text{at } \mathcal{V}, \tag{2.1a}$$

$$\boldsymbol{0} = -\boldsymbol{\nabla} p + \boldsymbol{\nabla} \cdot \boldsymbol{\tau} \qquad \text{at } \mathcal{V}, \tag{2.1b}$$

where $p$, $\boldsymbol{v}$ and $\boldsymbol{\tau} = \mu[\boldsymbol{\nabla}\boldsymbol{v} + (\boldsymbol{\nabla}\boldsymbol{v})^T]$ are the pressure, the velocity vector and the viscous stress tensor, respectively.

In the reference frame attached to the bubble moving at the equilibrium velocity $V$ in the axial direction, the velocity of the liquid at the wall writes

$$\boldsymbol{v}(\boldsymbol{x}) = -V\boldsymbol{e}_x \qquad \text{at } \Sigma_W, \tag{2.2}$$

whereas the equilibrium condition in the axial direction for the bubble writes

$$\int_{\Sigma_B} (-p\boldsymbol{n} + \boldsymbol{n} \cdot \boldsymbol{\tau}) \cdot \boldsymbol{e}_x \, \mathrm{d}\Sigma = 0, \tag{2.3}$$

where $\boldsymbol{n}$ and $\mathrm{d}\Sigma$ are the outer normal and the differential surface, respectively.

The velocity profile and pressure gradient are periodic in the axial direction with period $L$. Consequently, the pressure field between two different cross-sections differ by a constant pressure, and the boundary conditions are

$$p(\boldsymbol{x}+L\boldsymbol{e}_x) = p(\boldsymbol{x}) - \Delta p + L\partial_x p_P \qquad \text{at } \Sigma_{cross}, \tag{2.4a}$$

$$\boldsymbol{v}(\boldsymbol{x}+L\boldsymbol{e}_x) = \boldsymbol{v}(\boldsymbol{x}) \qquad \text{at } \Sigma_{cross}, \tag{2.4b}$$

$$\partial_x \boldsymbol{v}(\boldsymbol{x}+L\boldsymbol{e}_x) = \partial_x \boldsymbol{v}(\boldsymbol{x}) \qquad \text{at } \Sigma_{cross}, \tag{2.4c}$$

where $\Sigma_{cross}$ is any cross section and the pressure drop is due to the Poiseuille flow, i.e. $\partial_x p_P = -32\mu J/d_h^2$ in cylindrical microchannels, modified by the presence of the bubble, $\Delta p$. The superficial velocity $J$ is defined by

$$\int_{\Sigma_{cross}} (\boldsymbol{v} \cdot \boldsymbol{e}_x + V - J) \, \mathrm{d}\Sigma = 0. \tag{2.5}$$

Note that (2.4)-(2.5) must be imposed only at one arbitrary cross-section because of the periodicity. The solution is independent of the convenient choice of the position of $\Sigma_{cross} \equiv \Sigma_{OUT}$ and the center of the bubble $x = 0$.

The tangential stress balance and impermeability condition at the bubble surface write

$$-p\boldsymbol{n} + \boldsymbol{n} \cdot \boldsymbol{\tau} = -p_S \boldsymbol{n} \qquad \text{at } \Sigma_B, \tag{2.6a}$$

$$\boldsymbol{n} \cdot \boldsymbol{v} = 0 \qquad \text{at } \Sigma_B, \tag{2.6b}$$

where $\boldsymbol{n}$ is the outer normal and the surface variable $p_S$ is the impermeable surface pressure that the liquid exerts on the bubble surface. It should be noted that these boundary conditions also correspond to inviscid droplets.

### 2.2. Modelling the dissolution

The concentration of the gas dissolved in the liquid, $C(\boldsymbol{x},t)$, is governed by the unsteady advection-diffusion equation

$$\partial_t C + \boldsymbol{v} \cdot \boldsymbol{\nabla} C = D \nabla^2 C \qquad \text{at } \mathcal{V}, \tag{2.7}$$

together with the impermeability of the gas through the wall

$$-\boldsymbol{n} \cdot D \boldsymbol{\nabla} C = 0 \qquad \text{at } \Sigma_W, \tag{2.8}$$

the saturation concentration at the bubble surface

$$C = C_{sat} \qquad \text{at } \Sigma_B, \tag{2.9}$$

where we assume that $\Sigma_B$ is steady, and the periodicity conditions

$$C(\boldsymbol{x}) = C(\boldsymbol{x} + L\boldsymbol{e}_x) \qquad \text{and} \qquad \partial_x C(\boldsymbol{x}) = \partial_x C(\boldsymbol{x} + L\boldsymbol{e}_x) \qquad \text{at } \Sigma_{cross}. \tag{2.10}$$

Under this assumption, the equilibrium solution of the system (2.7)-(2.9) is $C(\boldsymbol{x}, t \to \infty) = C_{sat}$. In order to avoid transient analysis, it is convenient to apply separation of variables as

$$C(\boldsymbol{x},t) = C_{sat} + \sum_n^{\infty} \theta_n(\boldsymbol{x})[C_n(t) - C_{sat}], \tag{2.11}$$

where $\theta_n$'s are dissolution modes which decay as

$$\frac{\mathrm{d}}{\mathrm{d}t}(C_{sat} - C_n) = -\sigma_n(C_{sat} - C_n), \tag{2.12}$$

where $\sigma_n$ is the decay rate of the $n$-th mode. For illustration, considering only the first mode and no spatial dependencies, i.e. $\theta_{n>1} = 0$, if $\theta_1 = 0$ the concentration corresponds to the saturation concentration $C = C_{sat}$, whereas if $\theta_1 = 1$ the concentration corresponds to the concentration of that mode $C = C_1$, which is smaller than the concentration of saturation $C_{sat}$ in the case of dissolution. Thus an increase in $\theta$ corresponds to lower concentration, such as vanishing concentration corresponds to $\theta_1 = C_{sat}/(C_{sat}-C_1)$.

In effect, introducing (2.11) and (2.12) into (2.7)-(2.9), gives

$$-\sigma_n \theta_n + \boldsymbol{v} \cdot \boldsymbol{\nabla} \theta_n = D \nabla^2 \theta_n \qquad \text{at } \mathcal{V}, \tag{2.13}$$

together with the boundary conditions

$$\theta_n = 0 \qquad \text{at } \Sigma_B, \tag{2.14a}$$

$$\boldsymbol{n} \cdot \boldsymbol{\nabla} \theta_n = 0 \qquad \text{at } \Sigma_W, \tag{2.14b}$$

$$\theta_n(\boldsymbol{x}) = \theta_n(\boldsymbol{x} + L\boldsymbol{e}_x) \qquad \text{and} \qquad \boldsymbol{n} \cdot \boldsymbol{\nabla} \theta_n(\boldsymbol{x}) = \boldsymbol{n} \cdot \boldsymbol{\nabla} \theta_n(\boldsymbol{x} + L\boldsymbol{e}_x) \qquad \text{at } \Sigma_{cross}. \tag{2.14c}$$

The system of equations (2.13)-(2.14) sets an eigenvalue problem.

The mass transfer coefficient associated to the $n$-th mode, $k_{\ell n}$ is defined by

$$\Sigma_B k_{\ell n} = \int_{\Sigma_B} \boldsymbol{n} \cdot (-D \boldsymbol{\nabla} \theta_n) \, \mathrm{d}\Sigma. \tag{2.15}$$

Since, in this work, we mainly focus on the slowest eigenmode $\theta_1$, we normalised $\theta_1$ as

$$\int_{\mathcal{V}} (\theta_1 - 1) \, \mathrm{d}\mathcal{V} = 0, \tag{2.16}$$

in order to let $C_1$ be the averaged concentration due to the first mode. The contribution of the truncated modes are discussed in sec. 5. Finally, using the normalisation (2.16), the system (2.13)-(2.14) and the divergence theorem together with (2.1$a$) and (2.6$b$), the mass transfer coefficient (2.15) corresponding to the first mode becomes

$$k_\ell = \frac{\sigma_1 \mathcal{V}}{\Sigma_B}. \tag{2.17}$$

### 2.3. Scaling and numerics

The systems of equations for the hydrodynamics (2.1)-(2.6) and for the dissolution (2.13)-(2.14) can be made dimensionless with the hydraulic diameter of the channel $d_h$, the superficial velocity of the flow $J$ and the viscous stress $\mu J/d_h$ as characteristic length, velocity and pressure, respectively. In particular, the substitution in the aforementioned equations of

$$d_h \to 1, \quad J \to 1, \quad \mu \to 1, \quad \varepsilon \to \tilde{\varepsilon}, \quad d \to \tilde{d}, \quad L \to \tilde{L}, \quad D \to Pe^{-1}, \tag{2.18}$$

leads to the dimensionless formulation with the dimensionless numbers

$$Pe = \frac{Jd_h}{D}, \quad \tilde{\varepsilon} = \frac{\varepsilon}{d_h}, \quad \tilde{d} = \frac{d}{d_h}, \quad \tilde{d}_\Sigma = \frac{d_\Sigma}{d_h}, \quad \tilde{L} = \frac{L}{d_h}, \tag{2.19}$$

the dimensionless domain variables

$$\tilde{\boldsymbol{x}} = \frac{\boldsymbol{x}}{d_h}, \quad \tilde{\boldsymbol{v}} = \frac{\boldsymbol{v}}{J}, \quad \tilde{p} = \frac{pd_h}{\mu J}, \tag{2.20}$$

the dimensionless surface variable

$$\tilde{p}_S = \frac{p_S d_h}{\mu J}, \tag{2.21}$$

and dimensionless global variables

$$\tilde{\mathcal{V}} = \frac{\mathcal{V}}{J}, \quad \Delta \tilde{p} = \frac{\Delta p d_h}{\mu J}, \quad \tilde{\sigma} = \frac{\sigma d_h}{J}. \tag{2.22}$$

In addition, the Sherwood number, $Sh = (k_\ell d_\Sigma)/D$, is the dimensionless counterpart of the mass transfer coefficient, and is rewritten using (2.17) as

$$Sh = \frac{\sigma_1 \mathcal{V}}{\pi d_\Sigma D} = \frac{\tilde{\sigma}_1 \tilde{\mathcal{V}}}{\pi \tilde{d}_\Sigma} Pe. \tag{2.23}$$

Tildes are removed in the following for the sake of clarity. Then, we use superscript (d) for dimensional quantities when used.

It is worth noting that the model for heat transfer is analogue where liquid can warm up or cool down, in the case of which concentrations, gas diffusivity in the liquid and Sherwood number are substituted by temperature, thermal diffusivity and Nusselt number, respectively.

The dimensionless counterparts of the aforementioned system of equations are solved using the finite element method (FEM). The equations are implemented in weak form, taking advantage of the axisymmetry (when applicable), using the software COMSOL. Deformable bubbles are considered using the *Moving Mesh* module implemented in the Arbitrary Lagrangian-Eulerian method. In the numerical simulations, we cover the range for $Pe$ from $10^{-2}$ to $10^6$ limited by the accuracy and convergence of the computations.

## 3. Results

In this section, we study how the mass transfer is affected by the channel diameter, bubble diameter which is $d_\Sigma = d$ for spherical bubbles, the distance between bubbles, the diffusivity and the superficial velocity. First, we carry out a systematic study about the effects of the dimensionless numbers $d$, $L$ and, especially, $Pe$ on the $Sh$ for centered and non-deformable bubbles with stress-free interface.

In fig. 2, we schematise the influence of the $Pe$ number on the $Sh$ number including representative concentration patterns in the different regimes. We classify these regimes relative to two criteria. The first one is the importance of streamline diffusion versus convection gauged by the magnitude of $Pe$. For small $Pe$ numbers, the streamline diffusion is dominant or comparable to convection. For larger $Pe$ numbers, convection is dominant as compared to the streamline diffusion, which can be safely neglected. The second criteria is the interaction between bubbles, gauged by the comparison of the magnitude of $L^{(d)}$ and any other characteristic convection-diffusion length. If $L^{(d)}$ is smaller than the relevant convection-diffusion length, the $Sh$ number and the concentration pattern should depend on $L^{(d)}$.

In the small $Pe$ regimes, there exists a characteristic convection-diffusion length $\ell_{CD}^{(d)} \sim D^{(d)}/J^{(d)}$ over which convection and streamline diffusion are of the same order of magnitude. On the one hand, when this distance is larger than the distance between bubbles, $L^{(d)}/\ell_{CD}^{(d)} \sim Pe\, L^{(d)}/d_h^{(d)} \ll 1$, streamline diffusion among a distance $L^{(d)}$ dominates over convection and then the $Sh$ number and the concentration field become independent of $Pe$, i.e. convection is negligible. This regime is named as pure diffusion regime (PD). On the other hand, when the convection-diffusion length fits between two bubbles, $L^{(d)}/\ell_{CD}^{(d)} \sim Pe\, L^{(d)}/d_h^{(d)} \gtrsim 1$, the streamwise diffusion takes place among the distance $\ell_{CD}^{(d)}$, and $Sh$ increases with $Pe$ as $\ell_{CD}^{(d)}/d_h^{(d)} \sim Pe^{-1}$ decreases. This regime is named as convection-diffusion regime (CD). In sec. 3.1 we characterise these regimes.

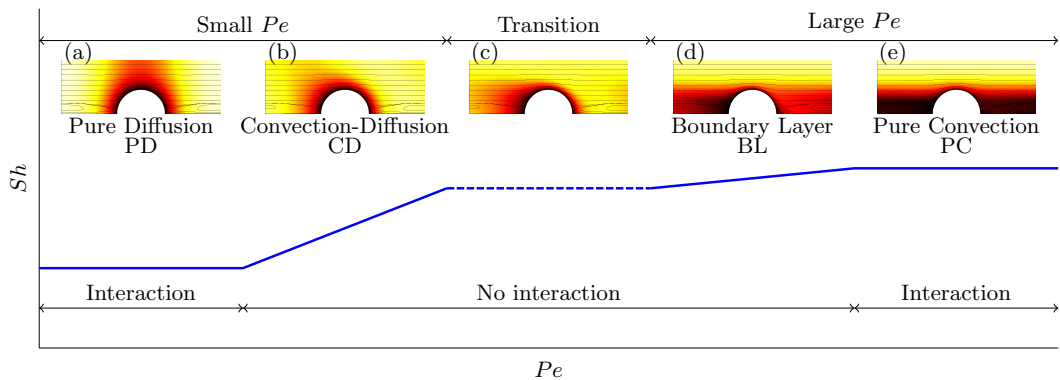

FIGURE 2. Schematic showing the influence of $Pe$ on $Sh$ and identifying the different dissolution regimes. These regimes are exemplified for the particular case of bubbles with stress-free interfaces, diameter $d = 0.45$, $L = 2$ and $Pe$ numbers (a) $Pe = 10^{-1}$, (b) $Pe = 10^1$, (c) $Pe = 10^2$, (d) $Pe = 10^3$ and (e) $Pe = 10^4$.

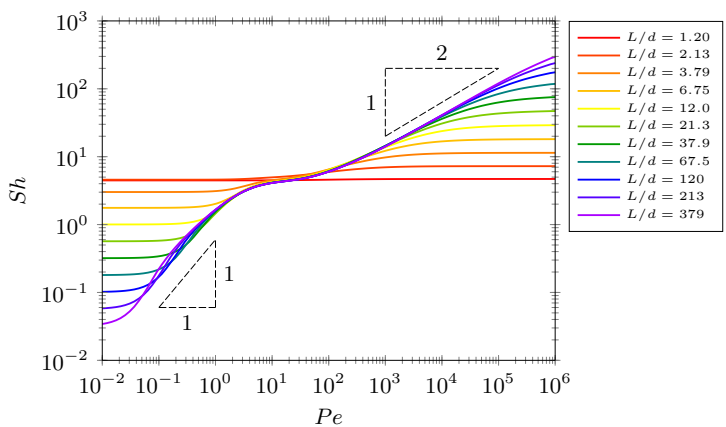

FIGURE 3. Influence of $Pe$ and $L$ on $Sh$ for bubbles of diameter $d = 0.45$.

In the large $Pe$ regimes, streamline diffusion is negligible as compared to convection and the crosswind diffusion becomes crucial. In effect, there appears a boundary layer around the bubble of width $\ell_{BL}^{(d)} \sim \sqrt{D^{(d)} d^{(d)} / J^{(d)}}$, where convection and crosswind diffusion are of the same order, together with a wake whose width is of the same order than the bubble diameter $d$. On the one hand, if the wake length is smaller than the distance between bubbles, the mass transfer takes place mainly in the boundary layer and we name this regime as boundary layer regime (BL). On the other hand, when the wake length, which increases with the $Pe$ number, becomes similar to the distance between bubbles $L^{(d)}$, the mass transfer through the wake becomes dominant as compared to the one through the boundary layer and both, concentration pattern and $Sh$ number become independent of the $Pe$ number. Contrarily to the BL regime, the concentration at the streamlines are not determined by the balance between convection and crosswind diffusion, and hence, remain unchanged. For this reason we name this regime as pure convection regime (PC). In sec. 3.2, we characterise these regimes. The regime between small and large $Pe$ regimes is referred to as the transition regime.

In fig. 3, we depict the influence of the $Pe$ number and the distance between bubbles $L$ on the $Sh$ for bubbles of diameter $d = 0.45$. We observe that, in effect, the $Sh$ strongly depends on the distance between bubbles in the 'interacting' regimes, PD and PC. In particular, we observe that the $Sh$ in the PD regime decreases with the length, since axial concentration gradients are smaller, whereas it increases in the PC regime because of the concomitant increase of the wake length. In the 'non-interacting' regimes, CD and BL, the characteristic length for the mass transfer is due to the convection-diffusion balance in the streamline and crosswind direction respectively and, hence, independent on $L$. It can also be observed that values of $Pe$ for which the crossover between 'interacting' and 'non-interacting' regimes takes place also depends on $L$. In particular, a decrease of $L$ shifts the crossover values of the $Pe$ towards the transition regime, eventually making all regimes to collapse into a single one as $L$ becomes sufficiently small. The system then behaves like a gas cylinder diffusing radially into an annular liquid (see sec. 3.3). As expected, a 1/1 slope emerges for the crossover between PD and CD regimes while a 1/2 slope emerges for the BL regime, in agreement with the theory of boundary layers. The $Sh$ number depends on $Pe$ as $Sh = 2 + kPe^\alpha$ in infinite medium (Clift 1978), where the dissolution is due to pure diffusion for small $Pe$ and through a boundary later for large $Pe$. These two different regimes are analogous to the PD and BL regimes in the present situation. Note that the BL regime does not depends on $L$ whereas the PD does. In the infinite medium, the PD regime is transient and the role of $L$ is played by the diffusion length $\sqrt{Dt}$. In the case of a confined train of bubbles the dependence of $Sh$ on $Pe$ is more intricate because of the additional regimes. In effect, the dependence shown in fig. 3 is a rational function.

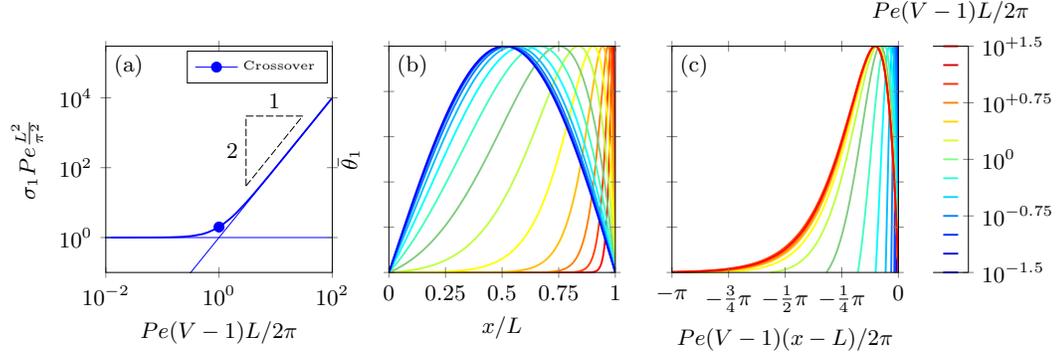

FIGURE 4. Transition between the PD and CD regimes of the 'axial model' (3.1). (a) Eigenvalues and (b-c) eigenmodes represented with two different scalings suitable for PD and CD, respectively.

### 3.1. *Small Pe*

In this subsection, we study both small $Pe$ number regimes, PD and CD. For this purpose, we can consider that if the distance between bubbles is much larger than the channel diameter, $L \gg 1$, the diffusion is mainly in the axial direction. Under this assumption, the concentration is uniform within any cross-section located at $x$ with exception at the vicinity of the bubbles. The dissolution modes are governed by the unidimensional counterpart of (2.13) in the axial direction and together with the boundary condition at the bubble surface (2.14a) and periodicity (2.14c). Note that (2.14b) is automatically fulfilled since the dissolution mode is considered as uniform in the cross-section. In addition, it is worth noting that the convection is imposed by the difference between superficial velocity and velocity of the bubble such that the relevant Péclet number becomes $Pe(V - 1)$. Consequently, if the bubble travels at the superficial velocity, the dissolution is purely diffusive. Then, the unidimensional model can be written as

$$-\bar{\sigma}_n Pe\bar{\theta}_n + Pe(V-1)\partial_x\bar{\theta}_n = \partial_{xx}\bar{\theta}_n \qquad \text{and} \qquad \bar{\theta}_n(0) = \bar{\theta}_n(L) = 0\,, \qquad (3.1)$$

where $\bar{\theta}_n(x)$ is the axial dissolution mode at the cross-section located at $x$ and $V$ is the velocity of the bubble obtained from the hydrodynamics. Although the periodicity (2.14c) is not imposed, it is automatically fulfilled because of the continuity of the solution. It is worth noting that the details of the concentration close to the bubble are not considered in this model. Eq. (3.1) is referred hereafter to as the 'axial model', as opposed to the 'full model', i.e. (2.13)-(2.14).

The analytical solution of the eigenvalue problem (3.1) writes

$$\bar{\sigma}_n Pe\frac{L^2}{\pi^2} = n^2 + \left[Pe(V-1)\frac{L}{2\pi}\right]^2, \qquad \bar{\theta}_n = e^{Pe(V-1)\frac{x}{2}}\sin\left(\frac{n\pi x}{L}\right)\,. \qquad (3.2)$$

It is depicted for $n = 1$ in fig. 4. For large $L$, the volume is approximately $\mathcal{V} \approx \pi L/4$ and the relation (2.23) between the $Sh$ and the eigenvalue $\sigma_1$ writes as $\frac{4}{\pi^2}ShLd \approx \sigma_1 Pe\frac{L^2}{\pi^2}$. Then, the dominant eigenvalue is depicted in fig. 4a where the crossover between the PD and CD regimes is marked with −•− and takes place for $Pe(V-1) = 2\pi/L$ and $\bar{\sigma}_1 Pe\frac{L^2}{\pi^2} = 2$. The eigenmode $\bar{\theta}_1$ is depicted in fig. 4b-c conveniently scaling the $x$ variable, with either $L$ or $\ell_{CD}$, to represent the eigenfunctions in the two regimes, PD or CD, respectively. In the PD regime, which takes place for $Pe(V-1) \ll 2\pi/L$, $Sh$ depends on the distance between bubbles as $Sh \sim L^{-1}$. The concentration pattern spans the distance between the bubbles as shown in fig. 4b. Contrarily, in the CD regime, which takes place in the other limit, $Pe(V-1) \gg 2\pi/L$, the diffusion is concentrated in a region of size $x \sim \ell_{CD} \sim Pe^{-1}$ as shown in fig. 4c.

In fig. 5, we compare the $Sh$ obtained from the 'axial model' and the 'full model'. For this purpose, in fig. 5a we conveniently rescale the data of fig. 3, i.e. $Pe$ and $Sh$ with $(V-1)L/2\pi$ and $4Ld/\pi^2$, respectively. In fig. 5a, we can observe that the transition in the small $Pe$ regime is recovered by the 'axial model' provided $L/d$ is sufficiently large to neglect the local effect close to the bubbles. In fig. 5b, we can observe that, again, the 'axial model' is recovered varying the bubble diameter, hence the bubble velocity. Indeed, the effect of the bubble diameter is considered through the bubble velocity as determined by solving the hydrodynamics, which is reported in our previous work Rivero-Rodriguez & Scheid (2018). In fig. 5c, we observe that the minimum value of $L$ to reproduce the 'axial model' depends on the $Pe$. We observe that local effects are more important for small values of $L$ and $d$, whose influence can be either to increase or decrease the dissolution.

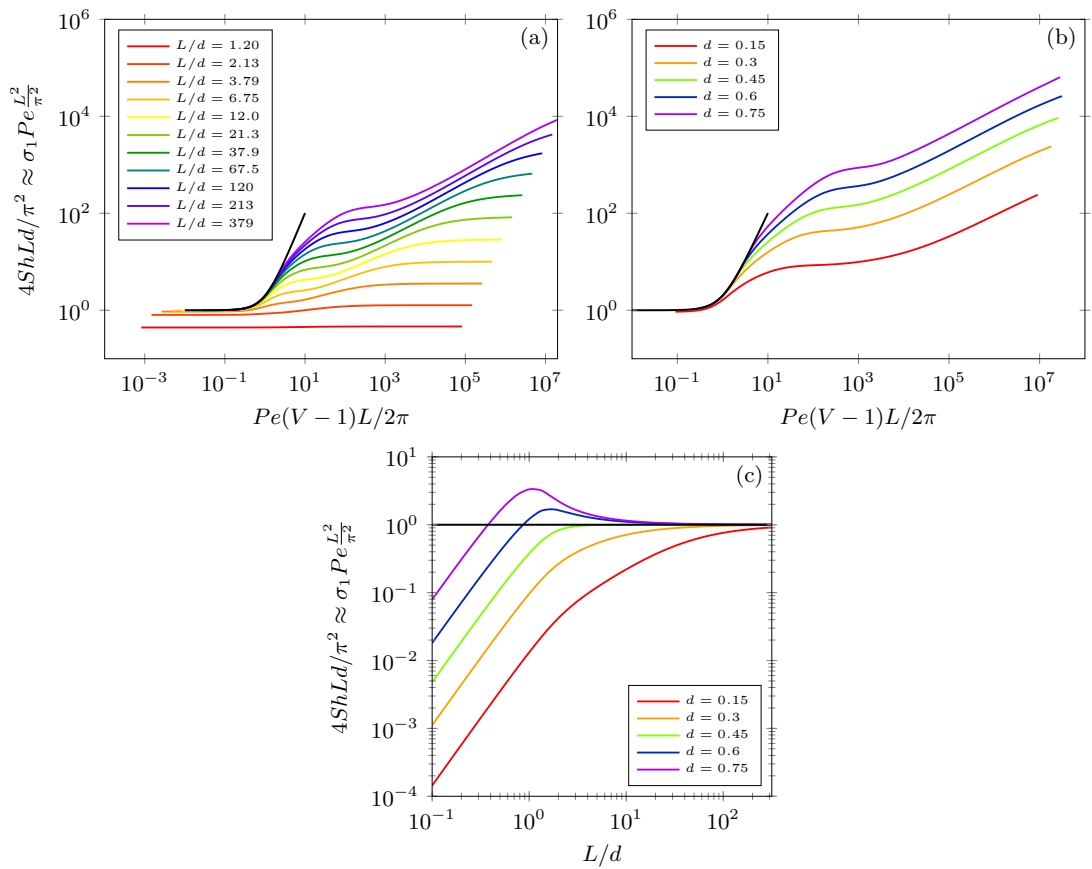

FIGURE 5. Transition between the PD and CD regimes of the 'full model' and —— 'axial model' (3.2) for (a) bubbles of diameter $d = 0.45$ and several channel lengths or (b) for channel length $L/d = 1.2 \times 10^{2.5}$ and several bubble diameters. (c) Influence of $d$ and $L$ on the PD regime.

### 3.2. *Large Pe*

In this subsection, we study the two large $Pe$ number regimes, i.e. BL and PC. In fig. 6, we can observe the influence of $Pe$, $d$ and $L$ on $Sh$ and the transition between both regimes. In fig. 6a, we depict the effect of the $Pe$ number and bubble size for a sufficiently large distance between bubbles, $L = 1.2 \cdot 10^{2.5}$, to have the transition of the PC regime at a larger $Pe$ such that $Pe = 10^6$ is in the BL regime. For this value of $L$, the BL regime is exhibited for values of $Pe$ over a few decades. We can observe a $1/2$ slope for the scaling of $Sh$ as function of $Pe$ similarly to the case of infinite medium (Haas *et al.* 1972) which differs from the work by Mikaelian *et al.* (2015a) who reported a slope of $2/5$. However, in this model we can observe that the effect of the confinement through the dependency on $d$ of $Sh$ is by the group $Pe\,d^3$ which emerges from the balance of convection and diffusion terms in the BL region. In effect, in shear flows where the velocity is of order $J$ and changes within distances of the order of $d_h$, such a balance writes

$$\frac{J^{(d)}}{d_h^{(d)}}\frac{d^{(d)}}{d_h^{(d)}} \sim \frac{D^{(d)}}{\left(\delta^{(d)}\right)^2}, \tag{3.3}$$

where changes in the velocity due to shear take place within the direction tangential to the bubble of size $d$, since in the normal direction there is no change of velocity due to the stress-free boundary conditions (2.6), and diffusion takes place within the boundary layer thickness $\delta^{(d)}$. Given that $k_\ell^{(d)} \sim D^{(d)}/\delta^{(d)}$, or equivalently $Sh \sim \left(\frac{\delta^{(d)}}{d^{(d)}}\right)^{-1}$, (3.3) can be rewritten as $Sh \sim \left(Pe\,d^3\right)^{1/2}$.

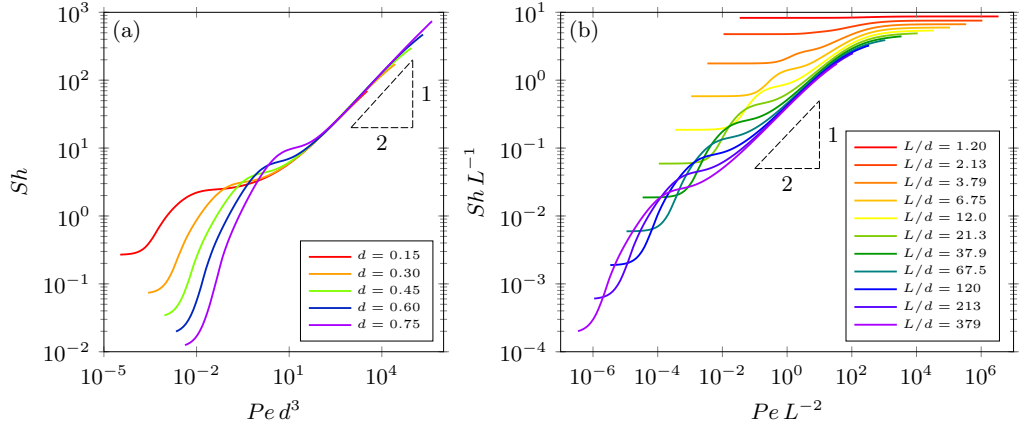

FIGURE 6. Influence of the $Pe$ number on the $Sh$ number in the large $Pe$ number regimes. (a) Scaling for the BL regime for a large channel $L = 1.2 \times 10^{2.5}$ and several bubble diameters and (b) transition to the PC regime for a bubble of diameter $d = 0.45$.

In fig. 6b, we conveniently rescale the data of fig. 3 taking into consideration the 1/2 slope in the BL regime, i.e. $Pe$ and $Sh$ rescaled with $L^{-2}$ and $L^{-1}$, respectively. It results into a collapse of the curves for sufficiently large $L$ where we can observe the influence of $L$ on the transition between the BL and the PC regimes. In particular, it can be observed that the dissolution becomes independent of the $Pe$ number and depends only on $L$ like $Sh \sim L$. It can be explained by the fact that the bubble mainly dissolves radially through the wake whose length is $L$ in the PC regime. Small disagreement might be due to local effects close to the bubble, i.e. thicker wakes in this region.

In effect, the concentration pattern in the PC regime is mainly uniform in the axial direction except in the vicinity of the bubble as shown in fig. 2e. In fig. 7a, we depict the concentration profile of the slowest mode $\theta_1$ in an arbitrary cross-section far from the bubble where the wake is uniform, $r$ being the cylindrical radial coordinate. We can observe that the core of the wake is saturated and diffusion is maximal at the contour of the wake of diameter $d_w$, decaying as getting closer to the wall where it vanishes. The radial counterpart of (2.13)-(2.14), corresponding to uniform velocity and concentration in the axial direction, can be written as

$$-\hat{\sigma}_n Pe\, \hat{\theta}_n = \frac{1}{r}\partial_r\left(r\partial_r\hat{\theta}_n\right) \quad \text{and} \quad \hat{\theta}_n\left(\frac{d_w}{2}\right) = \partial_r\hat{\theta}_n\left(\frac{1}{2}\right) = 0, \quad (3.4)$$

where the hat corresponds to variables of the model (3.4) referred to as the 'radial model'. The solution of this equation writes

$$\hat{\theta}_n(r) = Y_0\left(\sqrt{\hat{\sigma}_n Pe}\,\frac{d_w}{2}\right) J_0\left(\sqrt{\hat{\sigma}_n Pe}\,r\right) - J_0\left(\sqrt{\hat{\sigma}_n Pe}\,\frac{d_w}{2}\right) Y_0\left(\sqrt{\hat{\sigma}_n Pe}\,r\right), \quad (3.5)$$

where $\hat{\sigma}_n$ are obtained from the zeros of the following expression $J_0(\sqrt{\hat{\sigma}_n Pe}\,\frac{d_w}{2})Y_1(\sqrt{\hat{\sigma}_n Pe}\,\frac{1}{2}) - Y_0(\sqrt{\hat{\sigma}_n Pe}\,\frac{d_w}{2})J_1(\sqrt{\hat{\sigma}_n Pe}\,\frac{1}{2}) = 0$. Taking $d_w$ as the width of the recirculation region shown in fig. 10, the 'radial model', $\hat{\theta}_1$, reproduces the 'full model', $\theta_1$, as we show by dashed lines in fig. 7a. The value of $d_w$ depends on $d$ as shown in fig. 7b where we can see it is always smaller than $d$. It also corresponds to the size of the wake defined by the width of the recirculating regions shown by the streamlines in fig. 2e. In fig. 7c, we depict the relation between $\hat{\sigma}_1 Pe$ and $d_w$. In effect, the decay rate increases for larger bubbles as $d_w$ increases with $d$. Since $\sigma_1 Pe$ does not depend on $L$, the $Sh$ number, related to $\sigma_1 Pe$ by (2.23), is thus proportional to $L$ as chosen in the scaling of the PC regime shown in fig. 6b,

$$Sh(d,L) = \frac{L}{4d}\sigma_1[d_w(d)]Pe\,. \quad (3.6)$$

To summarise here, larger is the diameter, larger is the $Sh$ number.

### 3.3. *Dominating local effects. Small $L$*

In this subsection, we quantify the local effects which dominate for small distance between bubbles, i.e., $L \sim d$, already shown in fig. 5c for the PD regime. We observed in fig. 3 that the range of $Pe$ for 'non-interacting' regimes vanishes for sufficiently small $L$ depicting almost no influence of the $Pe$ number. Consequently, we can restrict to the PD regime without loss of generality. In fig. 8, we depict the slowest eigenvalue of (2.13)-(2.14) in the PD regime for several diameters and small values of $L$. In fig. 8a, we observe that for sufficiently separated bubbles the 'axial model', (3.2), is recovered (dashed line). However, when the size of the gap between bubbles $L - d$ is smaller than the bubble diameter, there is no influence of the distance between bubbles. It is so because the diffusion takes place mainly in the radial direction, as it happens in the dissolution of the wake in the PC regime, which leads to very similar $Sh$ in both regimes. The effect of the bubble diameter is depicted in fig. 8b, where we observe that $\sigma_1$ exponentially increases with $d$ as $\sigma_1 Pe = \exp(1.1966 + 3.8318d)$. In this situation, diffusion takes place in a distance that decreases as the bubble diameter increases, thus, $\sigma_1$ increases. The solution of the 'radial model' (3.4), with boundary conditions $\hat{\theta}_n(d/2) = \partial_r \hat{\theta}_n(\frac{1}{2}) = 0$, is depicted in dashed line. We observe that for small bubbles the diffusion is mainly radial since local effects are smaller than the distance between the bubble and the wall. However, for large bubbles local effects induce axial concentration gradients, increasing the diffusion length and decreasing the $Sh$ number with respect to the 'radial model' (3.4).

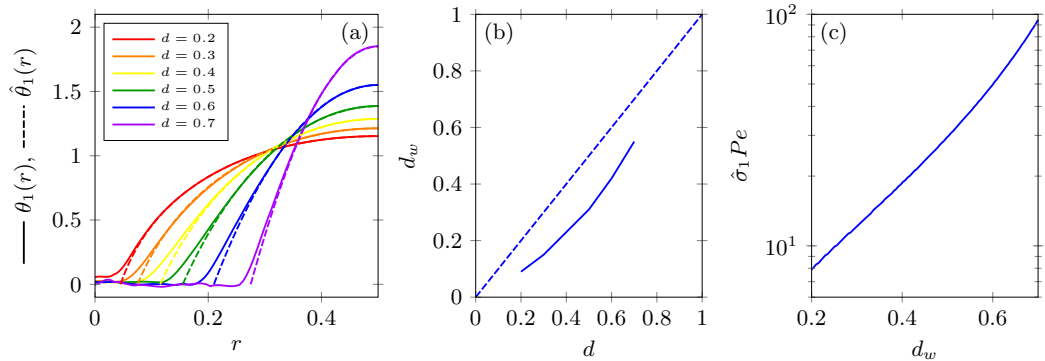

FIGURE 7. Radial diffusion in the PC regime for $L = 3$. (a) Concentration in a cross section far from the bubble for different bubble diameters (b) relation between the bubble diameter and the width of the wake and (c) influence of the diameter $d_w$ on the decay rate $\sigma_1$.

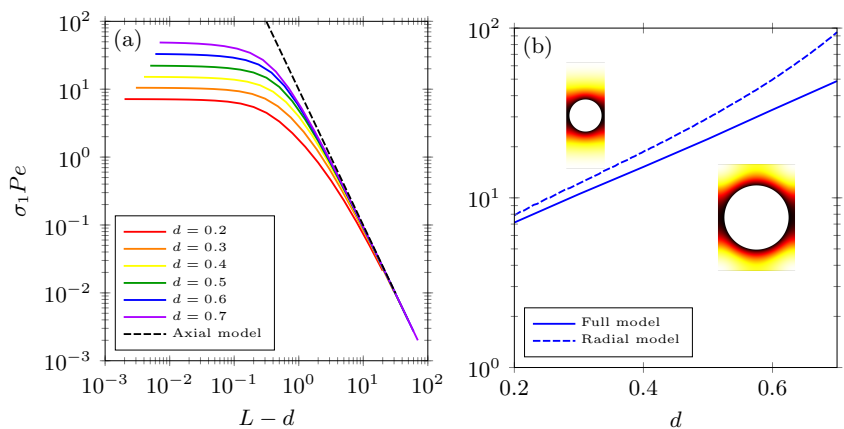

FIGURE 8. Local effect on the PD regime. Influence on the dissolution rate of (a) the distance between bubbles for several diameters $d$ and (b) of $d$ in the limit of small distances $L - d \ll 1$.

## 4. Model extensions

In this section, we consider some extensions of the hydrodynamic model described in sec. 2. We study the influence of a rigid interface, the deformability of the bubble, inertial forces and the off-centering of the bubble. The origin of rigid interface can be due to the presence of surfactants in the limit of rigid interface or substitution of the gas by either a high-viscosity liquid or a solid particle. For further details on the hydrodynamics of the model extensions, the reader is referred to our previous work Rivero-Rodriguez & Scheid (2018).

### 4.1. *Influence of a rigid interface*

Tailoring the surface rheology is known to modify the dissolution, either because of mechanical modification of the surface by adding surfactants Mikaelian *et al.* (2015*a*) or solid particles (Beltramo *et al.* 2017) or because of chemical modification of the mass transfer through the interface. In this subsection, we consider the effect of the former in the limit of spherical bubble with rigid interface. To this purpose, the hydrodynamics is governed by (2.1) together with (2.2), (2.3),(2.4), (2.5) and substituting (2.6) by the non-slip condition at the bubble surface

$$\boldsymbol{v}(\boldsymbol{x}) = \boldsymbol{0} \quad \text{at } \Sigma_B. \qquad (4.1)$$

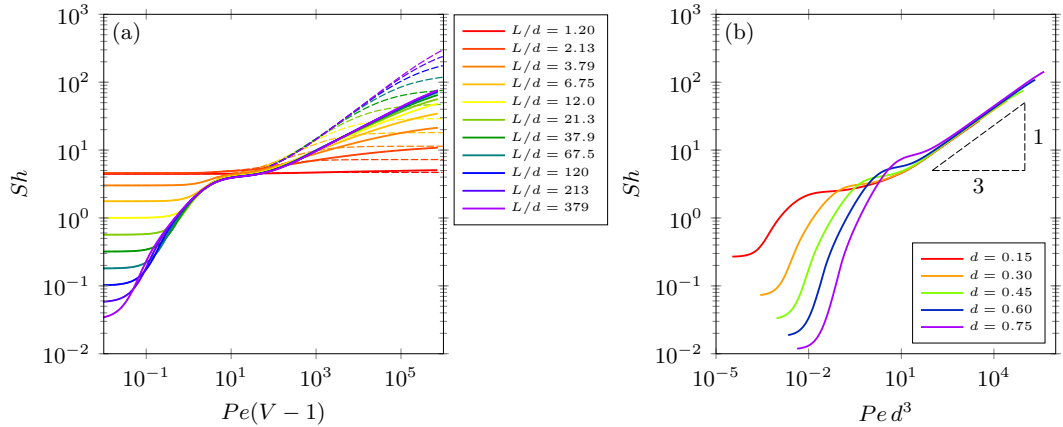

FIGURE 9. Influence of the $Pe$ number in the $Sh$ number for bubbles with rigid interface. (a) Comparison with respect to stress-free bubbles (dashed lines) for diameter $d = 0.45$ and several $L$. (b) Influence of the diameter for $L = 379.43d$.

It should be noted that these boundary conditions also correspond to very viscous droplets and solid particles. It is worth noting that (2.6b), which is used in the derivation of (2.17), is fulfilled, thus (2.17) also holds in this case.

In fig. 9, we depict the influence of $Pe$ on $Sh$. First, we compare it with the stress-free case in dashed lines in fig. 9a. We observe that the small $Pe$ regimes are not affected by the boundary conditions of the bubble surface, and thus, it is also in agreement with the 'axial model', which would also describe the PD and CD regimes for other boundary conditions such as those corresponding to viscous droplets or finite Marangoni stresses. Nevertheless, for large $Pe$ the mass transfer decreases with respect to the stress-free case due to smaller velocities close to the bubble surface and the influence of the boundary conditions have a strong influence on the dissolution. In particular, the slope is $1/3$ in the BL regime for large $L$ as shown in fig. 9b, as already reported by Mikaelian *et al.* (2015a) and in general agreement with mass transfer in shear flow (Clift 1978). It is worth noting that the interaction of bubbles reduces the slope in the BL regime and make the PC regime to disappear at least in the considered range of $Pe$. In fig. 9b, it can also be observed that in the BL regime, $Sh$ depends on $d$ proportionally. In effect, in shear flows where the velocity is of order $J$ and changes within distances of the order of $d_h$, such a balance writes

$$\frac{J^{(d)}}{d_h^{(d)}} \frac{\delta^{(d)}}{d_h^{(d)}} \sim \frac{D^{(d)}}{\left(\delta^{(d)}\right)^2} \qquad (4.2)$$

where changes in the velocity due to shear take place within the normal direction $\delta^{(d)}$ due to the rigid boundary condition (4.1), and diffusion takes place within the boundary layer thickness $\delta^{(d)}$. Given that $k_\ell^{(d)} \sim D^{(d)}/\delta^{(d)}$, or equivalently $Sh \sim \left(\frac{\delta^{(d)}}{d^{(d)}}\right)^{-1}$, (4.2) can be rewritten as $Sh \sim \left(Pe\, d^3\right)^{1/3}$.

In order to understand the interaction between bubbles it is convenient to depict the concentration patterns in the large $Pe$ regime. In fig. 10, we depict the concentration pattern of bubbles with $d = 0.5$, and distance between bubble $L = 2$ for different $Pe$ numbers. We can observe the strong influence of the flow pattern on the concentration one. It appears a mass transfer boundary layer with very narrow concentration wakes around the streamlines diverging from stagnation points or rings. When wakes extend up to the following bubble, the boundary layers are fed with mass coming from the previous bubble but the recirculation region is not as saturated as in the case of bubbles with stress-free interface as shown in fig. 10b-e. Consequently, the influence of interaction between bubbles is weaker and although it does not overcome the mass transfer through the boundary layer, it partially reduces the mass transfer by reducing the exponent of the power law. It is worth noting that in the case of bubbles with stress-free interfaces, the wakes are of a similar size as the bubble, hence having a stronger impact on the boundary layer. Additionally, the interaction between bubbles does not impose a concentration pattern independently of the $Pe$ number as it occurs in PC regime, at least for the $Pe$ numbers considered in the present work.

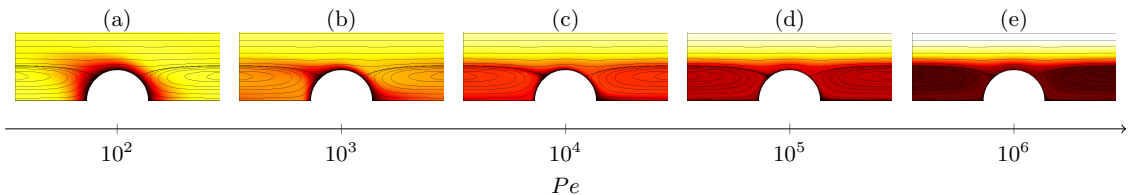

FIGURE 10. Streamlines and dissolution mode $\theta_1$ (dark red corresponds to saturation) in the case of bubbles with rigid interface and diameter $d = 0.45$ and distance between bubbles $L = 2$.

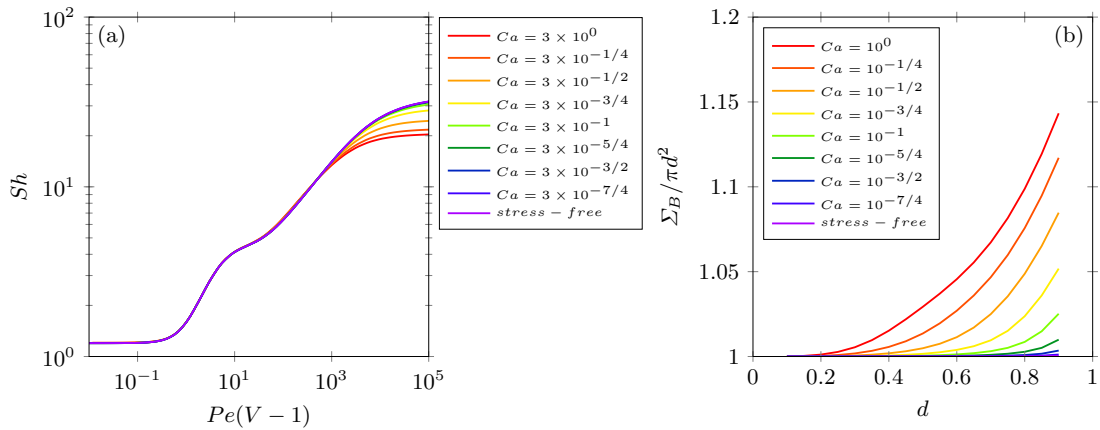

FIGURE 11. Influence of the deformability of bubbles, as gauged by $Ca$ for bubbles with diameter $d = 0.45$ and distance between bubbles $L = 12d$, on (a) the $Sh$ number and (b) the bubble surface.

### 4.2. *Influence of a deformable interface*

In this subsection, we consider the effect of deformable bubbles governed by surface tension. To this purpose, we consider the Young-Laplace equation for the stress balance in the normal direction. Thus, the hydrodynamics is governed by (2.1) together with (2.2), (2.4), (2.5) and substituting (2.3) and (2.6 $a$) by the equations for the bubble size and position, i.e.

$$\int_{\mathcal{V}_B} \mathrm{d}\mathcal{V} = \mathcal{V}_B \,, \tag{4.3a}$$

$$\int_{\mathcal{V}_B} \boldsymbol{x} \, \mathrm{d}\mathcal{V} = 0 \,. \tag{4.3b}$$

as well as the stress balance at the bubble surface

$$-p\boldsymbol{n} + \boldsymbol{n} \cdot \boldsymbol{\tau} + p_G \boldsymbol{n} = -\frac{1}{Ca}\boldsymbol{n}\boldsymbol{\nabla} \cdot \boldsymbol{n} \qquad \text{at } \Sigma_B \,, \tag{4.4}$$

where $p_G$ is the pressure of the gas. In the dimensionless counterpart of (4.4), $Ca = \mu J/\gamma$ is the capillary number. This system of equations is solved using the Boundary Arbitrary Lagrangian-Eulerian method developed by the authors Rivero-Rodriguez & Scheid (-) for the solution of PDEs on deformable domains.

In fig. 11, we plot the influence of the deformability of the bubble gauged by the $Ca$ number. In fig. 11a, the influence of $Ca$ number for the different regimes is depicted. Conveniently scaling the $Pe$ number by the relative flow velocity $V - 1$. We observe that the $Sh$ number does not depend on $Ca$ except in the PC regime. It is worth noting that the surface of the bubble $\Sigma_B$ and its equivalent diameter $d_\Sigma$, on which the mass transfer coefficient and $Sh$ numbers are based on, as well as the velocity of the bubble significantly depend on the $Ca$ number and volume of the bubble, $\mathcal{V}_B = \pi d^3/6$, as depicted in fig. 11b, respectively. The bubble surface increases (see fig. 11b) as well as the velocity of the bubble which can even increase above the maximum velocity of the Poiseuille flow for sufficiently large $Ca$ as already shown in our previous work Rivero-Rodriguez & Scheid (2018), i.e. $V > 2$. It is worth noting that the deformable case reduces, for $Ca \to 0$, to the bubble with stress-free spherical interface.

To understand the influence of $Ca$ in the PC regime, we depict the concentration pattern in this regime in fig. 12. We can observe how bubbles and wakes stretch as $Ca$ increases. Thus, since $Sh$ diminishes in the PC regime as wakes stretches as described in sec. 3.2, $Sh$ diminishes as $Ca$ increases.

It has to be noted that the extension including an elastic membrane at the bubble surface in the bubble dissolution is straightforward. Indeed, to consider the membrane, the boundary conditions (2.6) must be substituted by the volume and position constraints (4.3), together with the stress-balance considering the membrane and the no-slip conditions

$$-p\boldsymbol{n} + \boldsymbol{n}\cdot\boldsymbol{\tau} = -p_G\boldsymbol{n} + \frac{1}{Ca_m}\left(\nabla_S - \boldsymbol{n}\nabla_S\cdot\boldsymbol{n}\right)\cdot\left(\mathcal{I}\frac{2\nu}{1+\nu}\nabla_S\cdot\mathbf{u} + \nabla_S\mathbf{u}\right) \qquad \text{at } \Sigma_B, \qquad (4.5a)$$

$$\boldsymbol{v} = \boldsymbol{0} \qquad \text{at } \Sigma_B, \qquad (4.5b)$$

where $\mathbf{u}$ is the displacement of the membrane with respect to its equilibrium position and $Ca_m = \frac{\mu J}{E}$, being $E$ the elasticity of the membrane in $N/m$, is the capillary number based on the membrane elasticity and $\nu$ is the Poisson ratio.

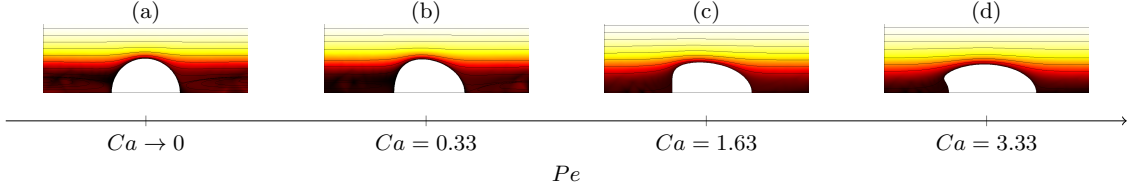

FIGURE 12. Streamlines and dissolution mode $\theta_1$ (dark red corresponds to saturation) for deformable bubbles with $d = 0.5$, $L = 1.5$ and in the PC regime, with $Pe = 10^5$.

From the numerical simulations, we have obtained that the $Sh$ is independent on $Ca_m$ and $\nu$ and reproduce the case of spherical bubbles with rigid interface. It is in is in accordance with the case of deformable bubbles, in which the deformability only affects the PC regime which does not appear in the case of rigid interfaces. Nevertheless, it is worth remembering that the $Sh$ number is based on $d_w$ which does depend on $Ca$, or $Ca_m$ and $\nu$, respectively.

An example of particular interest is the influence of the stiffness, or equivalently the deformability, of red blood cells (RBCs) on $O_2$ transfer rates. Storage of red blood cells (RBCs) increases their effective stiffness according to the literature Xu *et al.* (2018). Thus, the storage of RBCs modifies the $O_2$ transfer rates depending on the regime.

### 4.3. *Influence of inertia*

In this subsection, we study the effect of inertia in the bubble dissolution. To consider inertia, the hydrodynamics is governed by the Navier-Stokes equations, instead of the creeping flow (2.1),

$$\boldsymbol{\nabla}\cdot\boldsymbol{v} = 0 \qquad \text{at } \mathcal{V}, \qquad (4.6a)$$

$$Re\boldsymbol{v}\cdot\boldsymbol{\nabla}\boldsymbol{v} = -\boldsymbol{\nabla}p + \boldsymbol{\nabla}\cdot\boldsymbol{\tau} \qquad \text{at } \mathcal{V}, \qquad (4.6b)$$

where $Re = \rho J d_h/\mu$ is the Reynolds number, together with (2.2), (2.3), (2.4), (2.5) and (2.6).

After simulations for $Re$ up to 100, we observed that the changes in the hydrodynamics has no influence on $Sh$ within a 1% error. It is due to the fact that for small $Pe$ the system behaviour is well described by the 'axial model' which only consider the effect of the hydrodynamics through the velocity of the bubble. In the large $Pe$ regimes, the Schmidt number $Sc = \mu/(D\rho)$ are much larger than unity, $Sc \gg 1$. Thus, viscous boundary layer is thicker than the thermal one and, hence, the details of the viscous boundary layers tuned by the $Re$ do not affect the thermal one.

Thus, the influence of inertia is through the inertial migration force, responsible of off-center position of the bubbles. This effect was first reported by Segré & Silberberg (1962). In the next subsection, we consider the effect of off-centered positions of the bubble.

### 4.4. *Effect of off-centered positions*

In our recent work Rivero-Rodriguez & Scheid (2018), the equilibrium positions of deformable bubbles and rigid particles in inertial flows in the presence of an outer uniform force as function of the diameter, the $Re$ and the $Ca$ numbers are reported. In this subsection, we consider the effect of off-centered positions of the bubble, $\varepsilon$, measured from the center of the microchannel. In the case of stress-free bubbles the governing equations remain the same. However, in the case of rigid sphere in an off-centered position, the velocity at the bubble interface (4.1) is modified as

$$\boldsymbol{v}(\boldsymbol{x}) = \boldsymbol{\Omega}\times(\boldsymbol{x}-\boldsymbol{\varepsilon}) \qquad \text{at } \Sigma_B, \qquad (4.7a)$$

$$\boldsymbol{0} = \int_{\Sigma_B}\boldsymbol{n}\cdot\boldsymbol{\tau}\times(\boldsymbol{x}-\boldsymbol{\varepsilon})\,\mathrm{d}\Sigma, \qquad (4.7b)$$

where $\Omega$ is the rotational velocity of the bubble which results in moment equilibrium and $\boldsymbol{\varepsilon} = \varepsilon\boldsymbol{e}_y$.

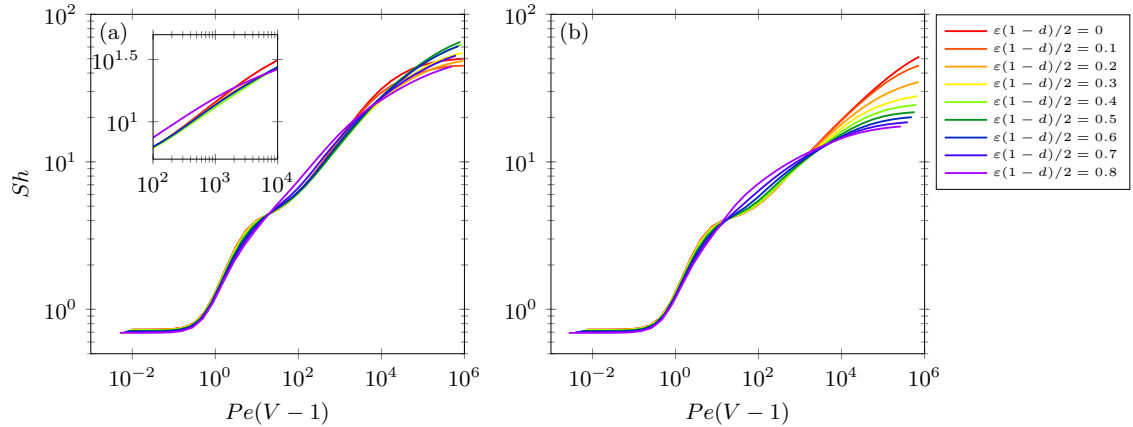

FIGURE 13. Effect of the off-center position for bubbles with diameter $d = 0.45$ and distance between bubbles $L = 20d$. (a) stress-free and (b) rigid surface.

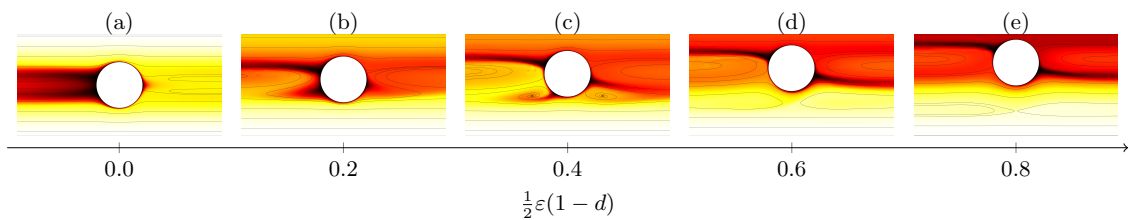

FIGURE 14. Streamlines and dissolution mode $\theta_1$ (dark red corresponds to saturation) in the case of bubbles with stress-free interface, diameter $d = 0.45$, distance between bubbles $L = 20d$ and $Pe = 10^4$.

In fig. 13, we plot the influence of off-centered positions for bubbles with both stress-free and rigid interfaces. On the one hand, we can observe that in the small $Pe$ regimes, the $Sh$ is affected by the eccentricity through the velocity of the bubble. On the other hand, the velocity field slightly affects the $Sh$ in the large $Pe$ regime. In particular, the effect on bubbles with stress-free interfaces is non-monotonic. Whereas for bubbles with rigid interface is monotonic, it tends to decrease the slope and it even reveals the PC regime for large $Pe$ in the considered range of $Pe$.

In fig. 14, we can observe the influence of the eccentricity $\varepsilon$ on the flow structure and hence on the dissolution modes, which we depicted in the region on large $Pe$ number, $Pe = 10^4$. We can observe how the stagnation regions, either points or lines, detach from the bubble surface. Observe that the divergent stagnation points/lines extract concentration from the boundary layer at the bubble surface. This intricate effect on the hydrodynamics might explain the non-monotonic of the dissolution, which makes it difficult to be quantified by analytical means.

## 5. Discussion on the assumptions and mode truncation

In real situations, the train of bubbles is formed at a certain location and, therefore, periodicity turns out not to be found in real systems. An example can be found in the work by Cubaud & Ho (2004), who report the formation of bubbles from a needle inside a microchannels. Once the bubble is formed, it is convected downstream at the same time it dissolves. For this reason, the concentration of gas in disperse phase increases downstream at the same time the bubble diameter decreases. It can also be observed that changes in the diameter of its own order of magnitude occurs when the bubble has traveled several distance between bubbles. It visually reinforces that the assumption of quasi-steadiness and the periodicity of the hydrodynamics practically holds. However, no arguments based on visual inspection can infer that the periodicity of concentration practically holds. Concerning the truncation of the modal expansion, it is worth mentioning that it fails for the initial stages when also the formation of the bubble and the presence of the needle leads to a failure of the assumptions.

In the next subsections, we discuss and quantify the accuracy of the truncation at the first eigenmode, namely the slowest one, and the two main assumptions: the quasi-steadiness assumption and periodicity.

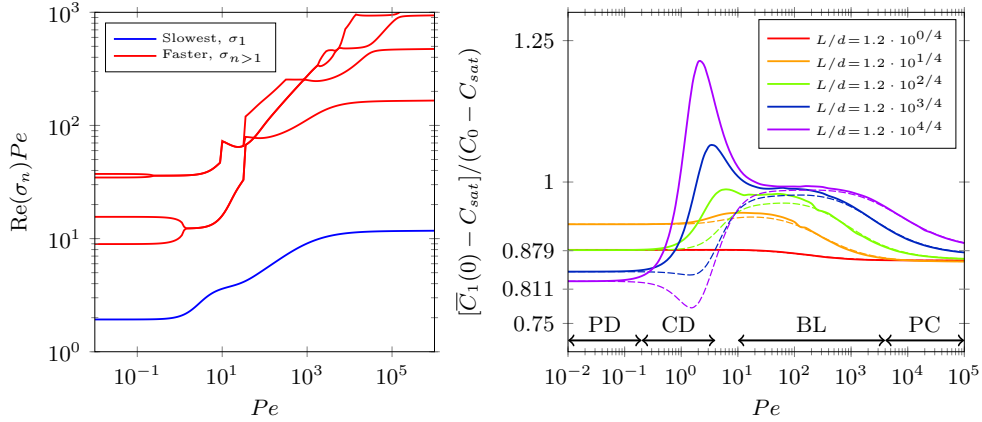

FIGURE 15. Influence of faster dissolution modes for bubbles with diameter $d = 0.45$ and stress-free interfaces. (a) Decay rates of dissolution modes for bubbles separated by a distance $L = 5d$. (b) Contribution of the first mode to a uniform initial concentration field for bubbles. Dashed lines corresponds to $[\overline{C}_{1,H}(0) - C_{sat}]/(C_0 - C_{sat})$. Values from the axial model (PD) and radial model (PC) are included.

### 5.1. *Modal expansion truncation*

To quantify the effect of neglecting faster modes, we depict in fig. 15 the real part of the decay rate $\sigma_n$ corresponding to the slowest five modes which might be oscillatory. It can be observed that the slowest mode is the considered one for the whole range of $Pe$ and that the difference with faster modes increases for increasing $Pe$. By neglecting the fast modes, it is implicitly assumed that they decay instantaneously as compared to the slowest one. Thus it is convenient to understand which is the contribution of the eigenmodes to the initial concentration, being the slowest one of particular interest. To this purpose and for the sake of illustration, we will assume an uniform initial concentration field $C(\boldsymbol{x}, 0) = C_0$, which substituted in (2.11), leads to

$$C_0 - C_{sat} = \sum_n^\infty \left[C_n(0) - C_{sat}\right] \theta_n(\boldsymbol{x}), \quad (5.1)$$

where $C_n(0)$ can be obtained from the projection of (5.1) onto $\theta_m$

$$(C_0 - C_{sat}) \int_\mathcal{V} \theta_m \, d\mathcal{V} = \sum_n^\infty \left[C_n(0) - C_{sat}\right] \int_\mathcal{V} \theta_n \theta_m \, d\mathcal{V}, \quad \forall m. \quad (5.2)$$

Note that the system (2.13)-(2.14) is not Hermitian in general and $C_n(0)$ must be obtained from the fully coupled algebraic system of equations (5.2). If the differential operator is Hermitian, the eigenmodes are orthonormal, i.e. $\int_\mathcal{V} \theta_n \theta_m \, d\mathcal{V} = 0$, and in this case (5.2) is decoupled as

$$(C_0 - C_{sat}) \int_\mathcal{V} \theta_n \, d\mathcal{V} = \sum_n^\infty \left[C_{n,H}(0) - C_{sat}\right] \int_\mathcal{V} \theta_n \theta_n \, d\mathcal{V}, \quad \forall n. \quad (5.3)$$

where subindex $H$ refers to Hermitian.

On the one hand, once the initial modal amplitudes $C_n(0)$ are obtained from (5.2), their evolution in time can be obtained by integration of (2.12). Then, the mean concentration $\overline{C}(t)$ can be obtained by averaging (2.11) over the domain $\mathcal{V}$ as

$$\overline{C}(t) - C_{sat} = \sum_n^\infty \left[\overline{C}_n(t) - C_{sat}\right], \quad (5.4)$$

where the modal contributions have been identified as $\overline{C}_n(t) - C_{sat} = \frac{1}{\mathcal{V}} \left[C_n(0) - C_{sat}\right] \int_\mathcal{V} \theta_n \, d\mathcal{V}$. Note that $\overline{C}(0) = C_0$. On the other hand, if the initial amplitudes are obtained from (5.3), namely Hermitian initial amplitudes $C_{n,H}(0)$ their modal contributions, namely Hermitian modal contribution, can be written as

$$\overline{C}_{n,H}(0) = \frac{\left[\int_\mathcal{V} \theta_n \, d\mathcal{V}\right]^2}{\mathcal{V} \int_\mathcal{V} \theta_n^2 \, d\mathcal{V}}. \quad (5.5)$$

Note that, although the values $C_n(t)$ depend on the normalisation of the eigenfunctions $\theta_n$'s, the values of $\overline{C}_n(0)$ are independent of it.

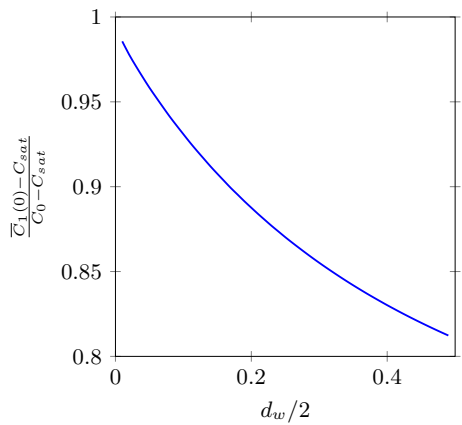

FIGURE 16. Modal contribution of the first mode to an uniform initial concentration field in the PC regime obtained from the 'radial model' (5.7).

In fig. 15b, we depict the modal contribution of the first mode to initial uniform concentration field, obtained for a sufficiently large basis to obtain results independently on its size, i.e. $n = 16$. It is worth noting that the deviation with respect to the unity represents the contribution of the rest of modes, which decay much faster than the slowest mode as shown in (15)a. These contributions significantly varies in the different regimes, gauged by the $Pe$ number and whose transition depends on $L$. Thus, in fig. 15b, the labels of the regimes are merely indicative.

We can observe that the contribution for the PD, BL and PC regimes does not depend on $Pe$, whereas it does for the CD regime for which it exhibits a peak. In the case of PD, an analytical expression for the modal contributions can be obtained using the 'axial model'. Indeed, after solving (5.2) for large $L$, where $\theta(\boldsymbol{x}) = \sin(n\pi x/L)$, the modal contributions in the PD regimes writes

$$\frac{\overline{C}_n(0) - C_{sat}}{C_0 - C_{sat}} \approx \frac{\left[\int_0^1 \sin(n\pi\xi)\,\mathrm{d}\xi\right]^2}{\int_0^1 \sin(n\pi\xi)^2\,\mathrm{d}\xi} = 4\frac{1-(-1)^n}{\pi^2 n^2}. \quad (5.6)$$

It vanishes for even values of $n$ and for odd ones it yields 0.8106, 0.0900, 0.0324, for $n = 1$, $n = 3$ and $n = 5$, respectively. In fig. 15b, it can be observed that the modal contribution of the first mode tends to 0.8106 as $L$ increases.

For the PC it reduces to a 0.89, whereas for the CD and transition regime it exhibits a peak at the same time it strongly differs from the Hermitian contribution. It is due the fact that convection competes with diffusion in the streamline direction. This feature could be of interest to enhance/control the mass transfer.

In the CD and transition regime, convection and diffusion are of the same order in the streamwise direction. It leads to the fact that modes are not orthonormal, i.e. $\int_{\mathcal{V}} \theta_n \theta_m \, \mathrm{d}\mathcal{V} \neq 0$. In this case, the modal contributions differ from the Hermitian modal contributions as shown in fig. 15b. This break of symmetry could be exploited to force the mass transfer in an efficient manner as in rectified diffusion (Hsieh & Plesset 1961).

In the BL regime, convection and diffusion occurs in orthogonal directions and the operator is Hermitian. In this regime, the mass transfer is through the boundary layer and the mean concentration is that outside the boundary layer in the first mode. Consequently, the contribution of the first mode is very close to unity.

Finally, in the PC regime, the diffusion and convection are in orthogonal directions. It leads to an Hermitian operator and the modal contributions can be obtained by the use of (5.5). Since the 'radial model' describes the dissolution in this regime, an analytical expression for the modal contributions can be obtained. It can be written, using (3.5), as

$$\frac{\overline{C}_n(0) - C_{sat}}{C_0 - C_{sat}} \approx \frac{\left[\int_{d_w/2}^{1/2} \hat{\theta}_n \, r\mathrm{d}r\right]^2}{\frac{1}{8}(1 - d_w^2) \int_{d_w/2}^{1/2} \hat{\theta}_n^2 \, r\mathrm{d}r}. \quad (5.7)$$

In fig. 16, we depict the contribution of the slowest dissolution mode obtained using (5.7). We can observe that the contribution of the first mode is larger for smaller bubbles and remains above 0.81 for bubble with a diameters up to the diameter of the channel.

We can conclude that the truncation is not valid for the very initial times but once the faster dissolution modes are extinguished, which always represents less than 10% of the total dissolution time, the dissolution is dominated by the slowest modes. Additionally, the faster dissolution modes do not necessarily contribute to the dissolution as the even pure diffusive modes and the modal contribution of the faster mode to the mean concentration is small as compared to the slowest mode.

### 5.2. *Quasi-steadiness assumption*

To obtain a criteria to quantify the validity of the quasi-steadiness assumption, it is necessary to compare characteristic dissolution and hydrodynamic times. The bubble dissolution time is of the order $t_{diss,B}^{(d)} \sim \mathcal{V}_B^{(d)} \rho_B / \left( \Sigma_B^{(d)} k_\ell^{(d)} |C| \right)$, where $\rho_B$ is the density of the bubble and $|C|$ is the dimensional concentration jump between the saturation and the domain averaged concentrations. The hydrodynamic time is of the order $t_{hyd}^{(d)} \sim d_h^{(d)} / J^{(d)}$. Then, the dissolution is quasi-steady if the dissolution time is larger than the hydrodynamic time (Plesset & Zwick 1954; Lohse *et al.* 2015). Using the definition of $Sh$ in (2.23), the equation for the validity of the quasi-steady assumption can be written as

$$\frac{t_{hyd}^{(d)}}{t_{diss,B}^{(d)}} \sim \frac{Sh(Pe)}{Pe} \frac{|C|}{\rho_B} \ll 1 \,. \tag{5.8}$$

It is worth noting that the quasi-steadiness assumptions is more valid for gases with smaller solubilities or for liquids with a certain amount of dissolved gas. Assuming ideal gas and Henry's law, the maximum value of $|C|/\rho_B$ for negligible surface tension, writes, considering that the minimum bulk concentration is zero,

$$\frac{|C|}{\rho_B} < \frac{C_{sat}}{\rho_B} = RTH \,, \tag{5.9}$$

where $R$ is the ideal gas constant, $T$ is the temperature and $H$ is the Henry constant. This value is around 1.2020 for $CO_2$ at a temperature $T = 25C$. Note this value is independent of the pressure of the bubble and the validity of the model decreases with the solubility of the gas and the temperature. The validity also depends on the $Pe$ number through the function $Sh(Pe)/Pe$. If the quasi-steadiness does not hold, the shrinking velocity of the bubble $\dot{R}$ must be taken into account. This can be formalised by

$$\boldsymbol{n} \cdot \boldsymbol{v} = -\dot{R} \qquad \text{at } \Sigma_B \,, \tag{5.10}$$

where $\dot{R}$ is the time derivative of the bubble radius, which will be referred to as shrinking velocity. Eq. (5.10) should then be used instead of (2.6*b*). It is possible to be done because in the Stokes equations there is no time derivative as in the Navier-Stokes equations which includes de effect of the inertia.

In fig. 17, we depict the influence of the shrinking velocity and the range of validity in terms of the $Pe$ number of the quasi-steadiness assumption. In fig. 17a we reproduce one of the cases already plotted in fig. 3 for $d = 0.45$ and $L = 12d$ for different shrinking velocities $\dot{R}$. It can be observed that PD and CD regimes are not affected by the shrinkage of the bubble, as also found in other diffusive configurations such as the dissolution of bubble clouds (Michelin *et al.* 2018). These are the regimes for which the quasi-steadiness assumption does not hold, as shown by the dashed lines for constant $Sh/Pe = \rho_B/|C|$. The regions above these lines represent the regions where the quasi-steadiness assumption does not hold. We also observe that the validity increases as either $Pe$ or $\rho_B/|C|$ increases. It is worth mentioning, that for liquids the factor $|C|/\rho_B$ is typically orders of magnitude smaller which reduces the range of validity, even though, the validity in the large $Pe$ regimes is still ensured for both bubbles and droplets. In fig. 17b, we depict the influence of the shrinking velocity on the velocity of the bubble. We can observe that the shrinkage of the bubble linearly increases its velocity in such a manner that the velocity of the rear front of the bubble is essentially the same. When the quasi-steadiness assumption is not valid, either the full transient behaviour must be solved or modal expansion should retain history effects (Michaelides 2003; Peñas-López *et al.* 2016).

Finally, despite the fact that we have referred in this paper to mass transfer, it is also applicable to heat transfer. In this case, the assumption of quasi-steadiness typically holds because changes of volume due to changes in temperature are much smaller than in the dissolution.

### 5.3. *Periodicity assumption*

There are two main sources of inaccuracy of the periodicity assumption. The first one refers to the edge effect of a finite train of bubbles. The second one refers to fast changes in the bulk concentration as compared to the hydrodynamic time.

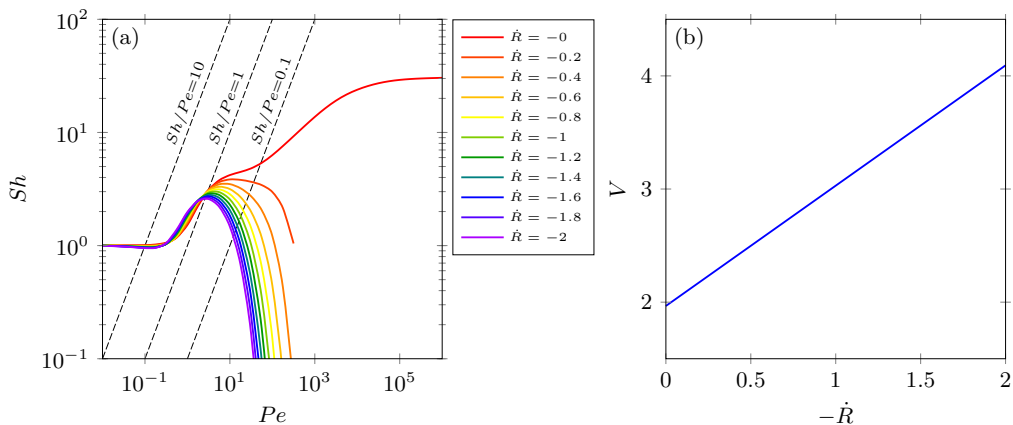

FIGURE 17. Validity of the quasi-steadiness assumption and influence of the shrinking velocity $\dot{R}$ for bubbles with diameter $d = 0.45$ and distance between bubbles $L = 12d$ on (a) the $Sh$ number and (b) the velocity of the bubble.

As fas as the edge effect is concerned, Michelin *et al.* (2018) report the dissolution of a stagnant cloud of bubbles. In their work, they explain how the bubbles within a layer close to the boundary of the clouds dissolve much faster than the inner ones. The thickness of this layer increases with time. This phenomena is referred to as the shield effect and has an analogue in this problem at the edge of the train where bubbles are completely disolved. In the work of Mikaelian *et al.* (2015b), a model is reported to obtain the concentration and bubble diameters at any downstream position of the bubbles. Although the phenomena is periodic, the average values of the previous variables are steady. Then, in this case the thickness of the edge effect is periodic with steady average value and spans to a few distance between bubbles. This effect is expected to be larger in 'interacting' regimes.

As it concerns the changes in the bulk concentrations, in the work Mikaelian *et al.* (2015b), it is reported that in the experiments by Cubaud & Ho (2004), there is an increase of concentration. However, if the increase of concentration is small as compared to the concentration jump $|C|$, it can be considered that the periodicity assumption practically holds. To obtain an estimate of this assumption, we have to compare the bulk dissolution time in which the bulk concentration changes $t_{diss,\mathcal{V}}^{(d)} \sim (C_{bulk,n}^{(d)} - C_{sat}^{(d)})/\dot{C}_n^{(d)}$ with the residence time $t_{hyd}^{(d)} \sim L^{(d)}/d_h^{(d)}$ which corresponds to the time in which a bubble travels the distance between bubbles, using (2.12) and (2.23),

$$\frac{t_{hyd}^{(d)}}{t_{diss,\mathcal{V}}^{(d)}} \sim \frac{Sh(Pe)}{Pe} d \ll 1, \quad (5.11)$$

which holds when the periodicity assumption is valid. In this case, sequent bubbles dissolve at the same rate in the same bulk concentration (2.11). On the contrary, if (5.11) does not hold, sequent bubbles dissolve at different rates due to different bulk concentration

$$C(\boldsymbol{x}, t) = C_{sat} + \sum_{n}^{\infty} \theta_n(\boldsymbol{x})[C_n(t) - C_{sat}] + \Delta C \, \boldsymbol{e}_x \cdot (\boldsymbol{x} - \boldsymbol{\varepsilon}), \quad (5.12)$$

which then results in an additional term in (2.14a), leading

$$\theta_n = -\Delta\theta \, \boldsymbol{e}_x \cdot (\boldsymbol{x} - \boldsymbol{\varepsilon}) \quad \text{at } \Sigma_B, \quad (5.13)$$

where $\Delta\theta = \Delta C/(C_{n,bulk} - C_{sat})$ and it must substitute (2.14a).

We then found that influences on the $Sh$ number are negligible, i.e. below 0.1%, for $\Delta\theta$ up to 10 while reasonable values are $\Delta\theta \lesssim 1$.

It is worth noting that Mikaelian *et al.* (2015a) includes a concentration drop in its formulation, i.e. in (2.10) similarly as for the pressure in (2.4a). In addition, they solve the steady advection-diffusion equation instead of the transient, which can be obtained by removing the temporal term in (2.7). The solution of the model of Mikaelian *et al.* (2015a) leads to an increment of concentration upstream, contrarily to the increment of concentration downstream reported in the experimental data by Cubaud & Ho (2004). Nevertheless, it does not compromise its validity in the range of $Pe$ and $L$ they studied in that work, which corresponds to a region within the BL regime, since the concentration drop turns out to be negligible. However, out of this range the concentration drop becomes significant. Furthermore, the $Sh$ number depends on the position of the arbitrary cross-section $\Sigma_{cross}$, revealing that this model is not appropriate for all the regimes. Contrarily, in the models considered in the present work, the results are independent of the choice of the arbitrary cross-section $\Sigma_{cross}$, what makes these models suitable for the whole range of $Pe$.

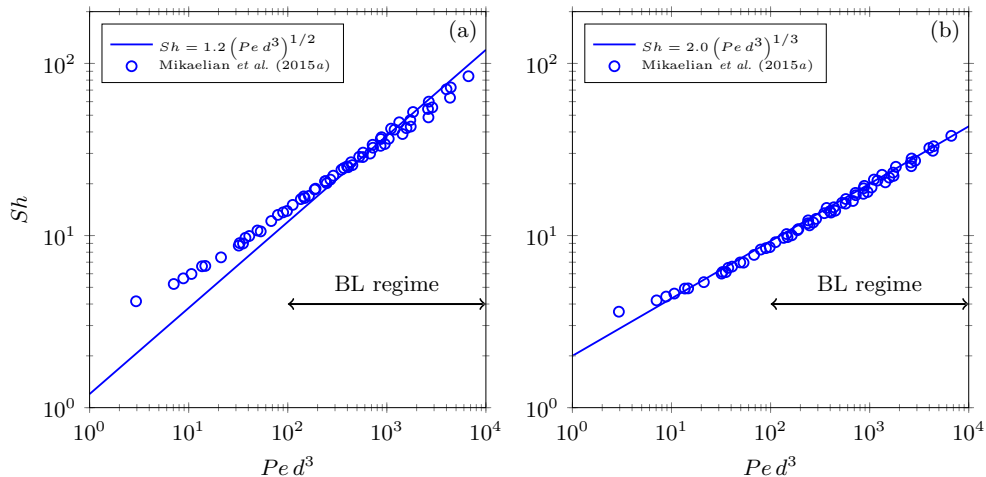

FIGURE 18. Comparison with Mikaelian *et al.* (2015a) in the BL regime for (a) stress-free and (b) rigid bubble surfaces.

In fig. 18, we compare our solution in the BL regime with the results by Mikaelian *et al.* (2015a). We observe a good agreement for large $Pe\,d^3$ where the bubble is in the BL regime, whereas the power law slightly overestimates the results of Mikaelian *et al.* (2015a) for smaller $Pe\,d^3$ in the transition regime, especially for stress-free boundary condition as expected outside the BL regime (see fig. 6a). It also explains the 2/5 power found by Mikaelian *et al.* (2015a) by fitting their data in a wider range than the BL regime for which the 1/2 power should rigorously apply in the case of bubbles with stress-free interface. The cross-over between the transition regime and the BL regime depends on $d$ and we have highlighted the region of $Pe\,d^3$ where the BL regime occurs independently of the value of $d$ within the considered range $0.15 < d < 0.75$.

After the discussion, we can conclude that the quasi-steadiness and periodicity assumptions are very appropriate for this problem because even if they do not hold, we have demonstrated that the results provided by the model are not affected by the relaxation of these assumptions.

## 6. Conclusions

In this work, the dynamics and dissolution of a train of bubbles in a liquid of constant properties is studied. We model the system under the assumption of quasi-steadiness of the bubble size and periodicity of the concentration field. We write the steady Stokes for the hydrodynamics and the eigenvalue counterpart of the unsteady advection-diffusion equation for the dissolution, obtained after separation of variables, and of which we compute the slowest mode. We consider periodic domains containing one bubble with impermeable microchannel wall both to liquid and gas. Concerning the bubble, we first consider stress-free boundary conditions at the surface of spherical and centered bubbles as the reference situation. Saturation concentration at the bubble surface is imposed. Then, expansions of the model are introduced such as rigid bubble surface, deformable bubble surface under finite surface tension, the effect of inertial forces and off-centered positions of the bubble.

We found five different regimes for which we provide a characterisation regarding two criteria. First, we regard at the relative importance between streamline diffusion and convection gauged by the Péclet number. Secondly we regard at the interaction between bubbles gauged by the distance between bubbles. On the one hand, for small Péclet numbers, convection is negligible or comparable to the streamline diffusion. If the convection-diffusion length is larger than the distance between bubbles, convection can be safely neglected and bubbles interaction is determinant in this purely diffusive regime, in which the Sherwood number does not depend on Péclet. Otherwise, bubbles do not interact and the mass transfer is constrained to a region of the size of the convection-diffusion length which decreases as Péclet number increases. This transition and the pure diffusive regime are characterised with the help of a simplified model which does not take into account details of the fluid flow but only the bubble velocity. On the other hand, for large Péclet numbers, the streamline diffusion is negligible as compared to the convection. In this situation, the convection-diffusion length dominates the mass transfer in the crosswind direction which is constrained to the boundary layer around the bubble and its wake. If this wake is shorter than the distance between bubbles, bubbles do not interact. Otherwise, the wakes of the preceding bubble modifies the boundary layer and convection becomes dominant and much larger than diffusion in the crosswind direction. In this regime, the streamlines represents iso-concentration lines, independently of the Péclet number, on which the dissolution does not depend anymore. In this regime, mass diffuses radially and another simplified model is developed to characterise this regime. Between the small and large Péclet regimes, there is a transition regime. To summarise, the regimes are pure diffusion, convection-diffusion, transition, boundary layer and pure convection, ordered for increasing Péclet number. Bubbles interact with each other in the first and last regimes of the previous list.

Table 1. Characterisation of the $Sh$ number as a function of $d$, $L$ and $Pe$ in the different regimes: pure diffusion (PD), convection-diffusion (CD), boundary layer (BL) and pure convection (PC).

| Regime | model | Stress-free interface | Rigid interface |
|---|---|---|---|
| PD&CD | 'axial' | $\frac{4}{\pi^2} Sh\, L\, d = 1 + \left[Pe(V-1)\frac{L}{2\pi}\right]^{1/2}$ | |
| BL | scaling | $Sh = 1.2\left(Pe\, d^3\right)^{1/2}$ | $Sh = 2.0\left(Pe\, d^3\right)^{1/3}$ |
| PC | 'radial' | $Sh = \frac{L}{4d}\sigma_1[d_w(d)]Pe$ | |

Next, we provide the scaling laws in the different regimes for the reference situation of spherical and centered bubbles with stress-free interface. In the pure diffusive regime, the Sherwood number is inversely proportional to the distance between bubbles and the diameter of the bubble while independent of the Péclet. The crossover between the pure diffusion and the convection-diffusion regimes occurs at a Péclet number, based on the difference of velocities between the bubble and the superficial velocity, and proportional to the distance between bubbles. The influence of the bubble size is through the velocity of the bubble which depends on its diameter. In the convection-diffusion regime, the Sherwood number is proportional to the distance between bubbles and the Péclet based on the difference of velocities between the bubble and the superficial velocity, but the decay ratio of the eigenmode does not. The transitional regime typically appears for Péclet numbers within the decade between $10^1$ and $10^2$. In the boundary layer regime, the Sherwood number scales with the Péclet number and the volume of the bubble with a $1/2$ exponent. For the largest Péclet number regime, the pure convection regime, the Sherwood number increases with the bubble diameter and is proportional to the distance between bubbles, which in this regime is also the length of the wake through which dissolution takes place.

We have also observed that for small distance between bubbles, the regimes for which bubbles do not interact with each other shrink until they vanish and the dissolution of both the pure diffusion and pure convection are very similar. We have characterised which is this distance regarding the pure diffusion regime and the influence of the diameter. In particular, the distance is of the order of the bubble size and the Sherwood number increases exponentially with the diameter.

As for the model expansions, the rigid interface does not affect the small Péclet number regimes, based on the difference of velocities, which depends on the bubble interface. In the boundary layer regime, the Sherwood scales with the Péclet number and the volume of the bubble with an exponent $1/3$. The boundary layer is thicker due to smaller convection of mass, related to smaller velocities which vanish at the bubble surface, as compared to the stress-free interface. The deformability of the bubble does not affect the Sherwood number, based on the diameter of the sphere with the same surface, for given Péclet number, based on the difference of velocities, in all the regimes with exception of the pure convection regime. In the latter regime, the stretching of the wake due to bubble deformation, decreases the bubble dissolution because gradients in the radial directions become smaller. However, the deformability of the bubble increases the bubble surface and the bubble velocity, thus increasing the bubble dissolution in any regime. The effect of the inertia on the Sherwood number is negligible for Reynolds numbers up to 100, which have been explored. However, it is well known that the balance between inertial and capillary migration forces and an outer force determines the off-centered position of the bubble and hence the bubble velocity, important for the small Péclet numbers regimes, and the flow structure around the bubble, important in the large Péclet numbers regimes. We have not found a systematic manner to quantify this effect but, although it is not negligible, it is rather small. Even though, the non-monotonic behaviour might be correlated to the detachment of stagnation points.

The characterisation of the $Sh$ number as a function of $d$, $L$ and $Pe$ number in the case of spherical centered bubbles with stress-free and rigid surface is summarised in table 1 for the various regimes.

Finally, we discuss the accuracy of the truncation of the modal expansion in the slowest mode and validity of the underlying hypotheses of the proposed model. Concerning the accuracy of the truncation of the eigenproblem for the dissolution, to consider only the slowest eigenmode reduces the validity at the early stages of the dissolution if the initial concentration field contributes to faster modes. In particular, the second mode damps almost one decade faster and the typical contribution of an initial constant concentration field is small as compare to the slowest eigenmode and may eventually vanish for the boundary layer regime. However, initial stages are also affected by the formation of the bubbles, whose influence on the dissolution we do not consider. Concerning the validity of the quasi-steadiness hypothesis, it strongly depends on the Péclet number, the density of the gas and the difference between the saturation concentration and the average concentration of gas in the liquid phase. However, the pure diffusion and convection-diffusion regimes, for which the quasi-steadiness hypothesis is not ensured, are not affected by the relaxation of this hypothesis. Concerning the validity of the periodicity hypothesis, it is not affected by its relaxation. Previous results in the literature where this hypothesis is not assumed are in reasonable agreement.

## Conflicts of interest

There are no conflicts to declare.